\documentstyle[11pt]{article}

\def\Term#1 #2 #3 /{\mbox{$\,^{#1}\!#2_{#3}$ }}
\def\Termo#1 #2 #3 /{\mbox{$\,^{#1}\!#2^o_{#3}$ }}
\def\be{\begin{equation}}
\def\ee{\end{equation}}

\begin{document}

\title{SOME NEW SIMPLIFICATIONS FOR GENERAL FORM OF ORBIT--ORBIT INTERACTION OPERATOR}
%\title{SIMPLE FORM OF ORBIT--ORBIT INTERACTION AND ITS` APPLICATION TO THE LIGHT ATOMS}
%\title{APPLICATION OF SIMPLE FORM OF ORBIT--ORBIT INTERACTION OPERATOR TO THE LIGHT ATOMS}
\author{G. Gaigalas \\
%EndAName
State Institute of Theoretical Physics and Astronomy, A. Go\v stauto 12, \\
2600 Vilnius, Lithuania}
\maketitle
\date{}

\begin{abstract}
The general tensorial form of the orbit--orbit interaction operator in the 
formalism of second quantization is presented. Such an expression is needed
to calculate both diagonal and off--diagonal matrix elements 
with respect to configurations, in a uniform way.
Some special cases are shown for which the orbit--orbit interaction
operator reduces to simple form. The importance of different contributions to the 
Breit--Pauli hamiltonian is investigated in ground states and ionization potentials 
for Li isoelectronic sequency in the systematic way.
\end{abstract}

\newpage

\section{Introduction}

For light atoms, there are two sorts of corrections to the non--relativistic 
energies
and wave functions that frequently are omitted but may need to be included
to improve the accuracy: the effect of the finite mass of the nucleus and
relativistic effects~\cite{Rudzika-book,book}. The lowest order corrections for the
former, which is particularly important for the very light atoms, can be
included through a redefinition of the Rydberg constant, $R_M = \frac{M}{M+m}
R_{\infty}$, for converting from atomic units to cm$^{-1}$, and a
mass-polarization correction given by the Hamiltonian
\begin{equation}
{\cal H}_{mp} = -\frac{1}{M} \sum_{i < J}\left({\bf p}_i\cdot {\bf p}_j\right),
\end{equation}
where $M$, $m$ are the mass of the nucleus and the mass of the electron,
respectively. Corrections for the relativistic effects can be obtained by
replacing the non--relativistic Hamiltonian by the Breit--Pauli Hamiltonian, 
${\cal H}_{BP}$, which includes the low--order terms of the
Dirac--Coulomb--Breit operator, namely terms of the order $\alpha ^{2}$
($\alpha$ is the fine structure constant).
This approach is called as Hartree--Fock--Pauli (HFP) approximation.

The Breit--Pauli Hamiltonian is often expressed in terms of operators $H_i$, $%
i=0,\ldots ,5$ introduced by Bethe and Salpeter~\cite{BS}, but it is also
constructive to separate the components according to their effect on the
spectrum as suggested by Glass and Hibbert~\cite{GH2}, namely
\begin{equation}
{\cal H}_{BP}={\cal H}_{NR}+{\cal H}_{RS}+{\cal H}_{FS},
\end{equation}
where ${\cal H}_{NR}$ is the ordinary non--relativistic many--electron
Hamiltonian. The {\em relativistic shift} operator ${\cal H}_{RS}$ commutes
with ${\bf L}$ and ${\bf S}$ and can be written
\begin{equation}
{\cal H}_{RS} = {\cal H}_{MC} + {\cal H}_{D1} + {\cal H}_{D2} + {\cal H}_{OO} +
{\cal H}_{SSC},
\end{equation}
where ${\cal H}_{MC}$ is the {\em mass correction} term
\begin{equation}
{\cal H}_{MC}=-\frac{\alpha ^2}8\sum_{i=1}^N{\bf p}_i^4 .
\end{equation}
The contact interactions describe the one-- and two--body {\em %
Darwin terms} ${\cal H}_{D1}$ and ${\cal H}_{D2}$. They are:
\begin{equation}
{\cal H}_{D1}=\frac{Z \alpha ^2 \pi}2\sum_{i=1}^N{\boldmath{\delta }}\left(
r_i\right) \qquad \mbox{and}\qquad {\cal H}_{D2} = 
-\pi \alpha ^2\sum_{i<j}^N{\boldmath{\delta }}\left( r_{ij}\right) .
\end{equation}
${\cal H}_{SSC}$ is the {\em spin--spin contact} term
\begin{equation}
{\cal H}_{SSC}=-\frac{8\pi \alpha ^2}3\sum_{i<j}^N\left( {\bf s}_i\cdot 
{\bf s}_j\right) \delta \left( {\bf r}_{ij}\right)
\end{equation}
and finally ${\cal H}_{OO}$ is the {\em orbit--orbit} term
\begin{equation}
\label{eq:oo}
{\cal H}_{OO}=-\frac{\alpha ^2}2\sum_{i<j}^N\left[ \frac{\left({\bf p}_i\cdot 
{\bf p}_j\right) }{r_{ij}}+
\frac{\left({\bf r}_{ij}\left( {\bf r}_{ij}\cdot {\bf p}_i\right)
{\bf p}_j\right) }{r_{ij}^3}\right] .
\end{equation}
One--particle operators ${\cal H}_{MC}$ and ${\cal H}_{D1}$ cause relativistic corrections
to the total energy. Two--particle operators ${\cal H}_{D2}$, ${\cal H}_{OO}$ and
${\cal H}_{SSC}$ define more precisely the energy of each term. The ${\cal H}_{MC}$,
${\cal H}_{D1}$, ${\cal H}_{D2}$ and ${\cal H}_{SSC}$ operators are included into the 
calculation by adding some terms to the radial integrals from non--realitivistic Hamiltonian.
The orbit--orbit operator needs separate calculations.

\medskip

The {\em fine--structure} operator ${\cal H}_{FS}$ describes interactions
between the spin and orbital angular momenta of the electrons, and does not
commute with ${\bf L}$ and ${\bf S}$ but only with the total angular
momentum ${\bf J}={\bf L}+{\bf S}$. So they describe the term splitting (fine structure). 
The fine--structure operator consists of three terms
\begin{equation}
\label{HamFS}{\cal H}_{FS}={\cal H}_{SO}+{\cal H}_{SOO}+{\cal H}_{SS}.
\end{equation}
The most important of these is the {\em spin--own orbit} interaction
${\cal H}_{SO}$ representing the interaction of the spin and angular magnetic
momentums of an electron in the field of the nucleus.
The {\em spin--other--orbit} ${\cal H}_{SOO}$ and {\em spin--spin} ${\cal H}_{SS}$
contributions may be viewed as corrections to the nuclear spin--orbit interaction 
due to the presence of other electrons in the system.
%Here ${\cal H}_{SO}$ is the {\em spin--own orbit},
%\begin{equation}
%\label{nuclSO}{\cal H}_{SO}=\frac{\alpha ^2Z}2\sum_{i=1}^N\frac 1{r_i^3}{\bf %
%l}_i\cdot {\bf s}_i
%\end{equation}
%${\cal H}_{SOO}$ is the {\em spin--other--orbit}
%\begin{equation}
%{\cal H}_{SOO}=-\frac{\alpha ^2}2\sum_{i<j}^N\frac{{\bf r}_{ij}\times {\bf p}%
%_i}{r_{ij}^3}\left( {\bf s}_i+2{\bf s}_j\right)
%\end{equation}
%and ${\cal H}_{SS}$ is the {\em spin--spin} terms.
%\begin{equation}
%{\cal H}_{SS}=\alpha ^2\sum_{i<j}^N\frac 1{r_{ij}}\left[ {\bf s}_i\cdot {\bf %
%s}_j-3\frac{\left( {\bf s}_i\cdot {\bf r}_{ij}\right) \left( {\bf s}_j\cdot
%{\bf r}_{ij}\right) }{r_{ij}^3}\right] .
%\end{equation}
%
%\medskip

The two--body terms 
${\cal H}_{SS}$ and ${\cal H}_{SOO}$ are complex leading to many radial
integrals. The complexity of the two--body ${\cal H}_{OO}$ operator however,
exceeds those of ${\cal H}_{FS}$, increasing the computer time required to
evaluate an interaction matrix. Thus, it has
been customary to omit the orbit--orbit effect from energy spectrum calculations.

\medskip

The first to treat the ${\cal H}_{OO}$ operator
in more detail were
Dagys {\sl et al.} \cite{DRVJ,DRKJ}, and Rudzikas {\sl et al.} \cite{RVJ}.
The expressions were obtained for the matrix elements of
$sp$, $sd$, $pp^{\prime }$ and $dd^{\prime }$ configurations cases, and for
the configurations with a single open shell. Later, the tensorial form
of this operator and the matrix elements between different configurations
were investigated by Beck \cite{B}, Match and Kern \cite{MK},
Walker \cite{W}, 
%Jones \cite{Jones}, 
Saxena {\sl et al.} \cite{SLF},
Dankwort \cite{Dan}, Anisimova and Semenov \cite{AS}, and
Anisimova {\sl et al.} \cite{AST}. Among them, Wybourne \cite{Wy}
had expressed the matrix elements of this operator inside one open shell
through the eigenvalues of Casimir operator. The irreducible tensorial form 
of the orbit--orbit operator was farther examined and
simplified by Kaniauskas and Rudzikas
\cite{KaniauskasR:73}. 

\medskip

The general method to
calculate the matrix elements of any two--body energy operator 
between arbitrary
configurations was proposed in Jucys and Savukynas \cite{JS},
Eissner {\sl et al.} \cite{E} (the latter is incorporated into
SUPERSTRUCTURE~\cite{E}), Glass~\cite{GH1}, Glass
and Hibbert~\cite{GH2}, and was investigated further by Badnell \cite{Badnell}.
An interesting recurrent relationship for the matrix element of orbit--orbit operator
between configurations with one open shell was obtained by
Ki\v ckin and Rudzikas \cite{KR}. Exhaustive tables of angular coefficients
for this operator are presented in Jucys and Savukynas \cite{JS}.
The matrix elements for configurations with
$f$--shells were investigated by Saxena and Malli \cite{SM69b}.

\medskip

Investigating the orbit--orbit interaction operator is made a lot easier after 
rewriting it in terms of
the products of tensorial operators and then applying the method of Racah.
So this paper aims to present the orbit--orbit operator in the style 
Gaigalas {\it et al.}~\cite{Gaigalas-meth5,Gaigalas-meth2} and 
using the integral properties find some new simplifications for the operator as well.
It enables us for evaluation of matrix elements of orbit--orbit interaction operator
i) to use full Racah algebra, namely allows to calculate both diagonal and 
off--diagonal matrix elements 
with respect to configurations, in a uniform way; use the unit tensors
in three spaces (orbit, spin, and quasispin), ii) to take into account new 
simplifications in general way. 

%\medskip

\section{Orbit--orbit interaction}

The tensorial form of {\it orbit--orbit} operator (\ref{eq:oo})
in general case is~\cite{Gaigalas-meth5}:
\begin{eqnarray}
\label{eq:b}
   {\cal H}^{oo}=\displaystyle
   \sum_k\left(
   H_{oo1}^{(kk0,000)}+H_{oo2}^{(kk0,000)}+H_{oo3}^{(kk0,000)}+H_{oo4}^{
   (kk0,00)}\right),
\end{eqnarray}
where
\begin{eqnarray}
\label{eq:c}
\lefteqn{
   H_{oo1}^{(kk0,000)}  }
   \nonumber \\[1ex]
   & &  =
   \frac{\alpha ^2}2k\left( k+1\right) \sqrt{2k+1}\left[ C_1^{\ (k)}\times
   C_2^{\ (k)}\right] ^{(0)}
   \nonumber \\[1ex]
   & &  \times
   \left\{ \frac 1{2k-1}~\frac{r_{<}^{k-1}}
   {r_{>}^k}-\frac 1{2k+3}~\frac{r_{<}^{k+1}}{r_{>}^{k+2}}\right\} \frac
\partial {\partial r_1}\frac \partial {\partial r_2},
\end{eqnarray}
\begin{eqnarray}
\label{eq:d}
\lefteqn{
   H_{oo2}^{(kk0,000)}  }
   \nonumber \\[1ex]
   & &  =
   \frac{i\alpha ^2}2\sqrt{k\left( k+1\right) \left( 2k+1\right) }\left[ \left[
   C_1^{\ (k)}\times L_1^{\ (1)}\right] ^{\left( k\right) }\times C_2^{\
   (k)}\right] ^{(0)}
   \nonumber \\[1ex]
   & & \times
   \left\{ -\frac{k-2}{2k-1}~\frac{r_{2}^{k-1}}{r_{1}^{k+1}}
            \epsilon (r_1-r_2)
           +\frac{k+1}{2k-1}~\frac{r_{1}^{k-1}}{r_{2}^{k}}
            \epsilon (r_2-r_1)
    +
            \frac{k}{2k+3}~\frac{r_{2}^{k+1}}{r_{1}^{k+3}}
            \epsilon (r_1-r_2) \right.
   \nonumber \\[1ex]
   & & 
   \left.
           - \frac{k+3}{2k+3}~\frac{r_{1}^{k+1}}{r_{2}^{k+2}}
            \epsilon (r_2-r_1)
   \right\} \frac 1{r_1}\frac
   \partial {\partial r_2}\left( 1+P_{12}\right),
\end{eqnarray}
\begin{eqnarray}
\label{eq:e}
\lefteqn{
   H_{oo3}^{(kk0,000)}  }
   \nonumber \\[1ex]
   & &  =
   -\alpha ^2\sqrt{2k+1}\frac{2k-1}{k+1}
   \nonumber \\[1ex]
   & &  \times
   \left[ \left[ C_1^{\ (k-1)}\times L_1^{\ (1)}\right] ^{(k)}\times \left[
   C_2^{\ (k-1)}\times L_2^{\ (1)}\right] ^{(k)}\right] ^{\left( 0\right) }%
   \frac{r_{<}^{k-1}}{r_{>}^{k+2}},
\end{eqnarray}
\begin{eqnarray}
\label{eq:f}
\lefteqn{
   H_{oo4}^{(kk0,00)} }
   \nonumber \\[1ex]
   & & =
   \frac{\alpha ^2}2\sqrt{2k+1}\left[ \left[ C_1^{\ (k)}\times L_1^{\
   (1)}\right] ^{(k)}\times \left[ C_2^{\ (k)}\times L_2^{\ (1)}\right]
   ^{(k)}\right] ^{\left( 0\right) }
   \nonumber \\   \nonumber \\[1ex]
   & & \times \left\{ \frac{\left( k-2\right)
   \left( k+1\right) }{2k-1}~\frac{r_{<}^{k-2}}{r_{>}^{k+1}}-\frac{k\left(
   k+3\right) }{2k+3}~\frac{r_{<}^k}{r_{>}^{k+3}}\right\} .
\end{eqnarray}
The $\epsilon (x)$ in (\ref{eq:d}) is a Heaviside step--function,
\begin{equation}
\label{eq:Tensorial-ss-e}
\epsilon (x)=\left\{
\begin{array}{ll}
1 ; & \mbox{ for } x>0, \\ 0 ; & \mbox{ for } x\leq 0.
\end{array}
\right.
\end{equation}
Here in equations (\ref{eq:b}) -- (\ref{eq:f}) we imply that a tensorial structure 
indexed by $(k_1 k_2 k, \sigma _1 \sigma _2 \sigma )$
at $H_{oo1}$, $H_{oo2}$, $H_{oo3}$ and $H_{oo4}$
has rank $k_1$ for electron 1, rank
$k_2$ for electron 2, and a resulting rank $k$ in the $l$ space, and corresponding
ranks $\sigma _1 \sigma _2 \sigma$ in the $s$  space. So four terms of 
orbit--orbit operator have the same tensorial structure (kk0 000), summed over 
$k$ in ${\cal H}^{oo}$ expression (\ref{eq:b}). It means that orbit--orbit 
operator is scalar in $s$ space.

\medskip

The general expression for any two--particle operator proposed
by Gaigalas {\it et al.}~\cite{Gaigalas-meth2} is sutable for evaluation of
diagonal and non--diagonal matrix elements in uniform way. 
It allows one to make the most of the advantages of Racah algebra
(see Racah \cite{Racaha,Racahb,Racahc,Racahd}). So further we will investigate
the orbit--orbit interaction in the framework of this formalism.
This expression has tensorial form:
\begin{eqnarray}
\label{eq:Tensorial-d}
\lefteqn{
   \widehat{G}^{\left( \kappa_1 \kappa_2 k, \sigma_1 \sigma_2 k \right)} }
   \nonumber \\[1ex]
   & & =
   \frac{1}{2} \displaystyle \sum_{iji^{\prime }j^{\prime
   }}\left( ij\left| g\right| i^{\prime }j^{\prime }\right)
   a_ia_ja_{j^{\prime }}^{\dagger }a_{i^{\prime }}^{\dagger },
   \nonumber \\[1ex]
   & & \sim
   \displaystyle {\sum_{\alpha}}
   \displaystyle {\sum_{\kappa _{12},\sigma _{12},\kappa
   _{12}^{\prime },\sigma _{12}^{\prime }}}\Theta \left( \Xi \right) \left\{
   A_{p,-p}^{\left( kk\right) }\left( n_\alpha \lambda _\alpha ,\Xi \right)
   \delta \left( u,1\right) \right. 
   \nonumber \\[1ex]
   & & +
   \displaystyle {\sum_{\beta}} \left[ B^{\left( \kappa _{12}\sigma
   _{12}\right) }\left( n_\alpha \lambda _\alpha ,\Xi \right) \times C^{\left(
   \kappa _{12}^{\prime }\sigma _{12}^{\prime }\right) }\left( n_\beta \lambda
   _\beta ,\Xi \right) \right] _{p,-p}^{\left( kk\right) }\delta \left(
   u,2\right) 
   \nonumber \\[1ex]
   & & +
\displaystyle {\sum_{\beta \gamma}}
\left[ \left[ D^{\left( l_\alpha s\right) }\times D^{\left( l_\beta
s\right) }\right] ^{\left( \kappa _{12}\sigma _{12}\right) }\times E^{\left(
\kappa _{12}^{\prime }\sigma _{12}^{\prime }\right) }\left( n_\gamma \lambda
_\gamma ,\Xi \right) \right] _{p,-p}^{\left( kk\right) }\delta \left(
u,3\right) 
   \nonumber \\[1ex]
   & & \left. +
\displaystyle {\sum_{\beta \gamma \delta}}
\left[ \left[ D^{\left( l_\alpha s\right) }\times D^{\left( l_\beta
s\right) }\right] ^{\left( \kappa _{12}\sigma _{12}\right) }\times \left[
D^{\left( l_\gamma s\right) }\times D^{\left( l_\delta s\right) }\right]
^{\left( \kappa _{12}^{\prime }\sigma _{12}^{\prime }\right) }\right]
_{p,-p}^{\left( kk\right) }\delta \left( u,4\right) \right\} ,
\end{eqnarray}
where
$\left( i,j|g|i^{\prime },j^{\prime }\right)$ is
the two--electron matrix element of operator
$G^{\left( \kappa_1 \kappa_2 k, \sigma_1 \sigma_2 k \right)}$,
$a _{i}$ is the electron creation,
$a_{j}^{\dagger}$ electron annihilation operators,
$iji^{\prime }j^{\prime }$, $i \equiv n_i l_i s_i m_{l_i} m_{s_i}$ and 
$\alpha$, $\beta$, $\gamma$, $\delta$ are strictly different.

\medskip

The summation in the (\ref{eq:Tensorial-d})
runs over the principle and the orbital quantum 
numbers of open shells. The first term represents
the case of a two--particle operator acting upon the same shell
$n_\alpha \lambda _\alpha$, the second term corresponds to operator
$\widehat{G}^{\left( \kappa_1 \kappa_2 k, \sigma_1 \sigma_2 k \right)}$
acting upon two different shells
$n_\alpha \lambda _\alpha$, $n_\beta \lambda _\beta$. When operator
$\widehat{G}^{\left( \kappa_1 \kappa_2 k, \sigma_1 \sigma_2 k \right)}$
acts upon three shells the third term in
(\ref{eq:Tensorial-d}) must be considered and when it acts upon four -- the
fourth one. 
We define in this expression the shells
$n_\alpha \lambda _\alpha$, $n_\beta \lambda _\beta$,
$n_\gamma \lambda _\gamma$, $n_\delta \lambda _\delta$
to be different.
In general case the number of combinations of $iji^{\prime }j^{\prime }$
(distributions) in (\ref{eq:Tensorial-d}) is infinite. 
In the work by Gaigalas {\em et al.} \cite{Gaigalas-meth2} an optimal 
number of distributions is chosen, which is enough for investigation of 
two--particle operator in general.

\medskip

The tensorial part of a two--particle operator in (\ref{eq:Tensorial-d}) 
is expressed in terms of operators of the type
$A^{\left( kk\right) }\left( n\lambda ,\Xi \right)$,
$B^{\left( kk\right) }(n\lambda ,\Xi )$,
$C^{\left( kk\right) }(n\lambda ,\Xi )$,
$D^{\left( ls\right) }$,
$E^{\left( kk\right) }(n\lambda ,\Xi )$
defined in \cite{Gaigalas-meth2}. They denote tensorial products
of those creation/annihilation operators that act upon a particular electron 
shell, $\lambda
\equiv ls$, and $u$ is the overall number of shells acted upon by a given
tensorial product of creation/annihilation operators. Parameter $\Xi $
implies the whole array of parameters (and sometimes an internal summation
over some of these is implied, as well) that connect $\Theta $
with tensorial products of creation/annihilation operators in the expression 
(\ref{eq:Tensorial-d})(see \cite{Gaigalas-meth2}).
These $\Theta \left( \Xi \right) $ are all
proportional to the submatrix element of a two--particle operator $g$,
\begin{equation}
\label{eq:m-ia}
\Theta \left( \Xi \right) \sim \left( n_i\lambda _in_j\lambda
_j\left\| g\right\| n_{i^{\prime }}\lambda _{i^{\prime }}n_{j^{\prime
}}\lambda _{j^{\prime }}\right) .
\end{equation}
So to obtain the general expression of orbit--orbit operator, analogous to
(\ref{eq:Tensorial-d}), the two--electron submatrix elements
(\ref{eq:m-ia}) must be defined.
In the following section we present these explicit expressions for this
operator.

\section{Submatrix elements for the orbit--orbit operator}

The sum of submatrix elements of three terms
$H_{oo1}^{(kk0,000)}$, $H_{oo2}^{(kk0,000)}$ and
$H_{oo4}^{(kk0,000)}$ is equal to
(see Badnell~\cite{Badnell}):
\begin{eqnarray}
\label{eq:Tensorial-oo-b}
\lefteqn{
   \left( n_i\lambda _in_j\lambda _j\left\|
   H_{oo1}^{(kk0,000)} + H_{oo2}^{(kk0,000)} +
   H_{oo4}^{(kk0,000)}
   \right\| n_{i^{\prime }}\lambda _{i^{\prime
   }}n_{j^{\prime }}\lambda _{j^{\prime }}\right) }
   \nonumber \\[1ex]
   & & =
   - 2 [k]^{1/2}
   \left( l_i\left\| C^{\left( k \right) }\right\| l_{i^{\prime }}\right)
   \left( l_j\left\| C^{\left( k \right) }\right\| l_{j^{\prime }}\right)
   \left( 1 - \delta \left( k,0 \right) \right)
   Z_k\left(
   n_il_in_jl_j,n_{i^{\prime }}l_{i^{\prime }}n_{j^{\prime }}l_{j^{\prime
   }}\right) ,
\end{eqnarray}
where we have used the conventional shorthand
notation $[k,...] \equiv (2k+1)\cdot ...$ and 
where
\begin{eqnarray}
\label{eq:i-a}
\lefteqn{
   Z_k\left( n_il_in_jl_j,n_{i^{\prime }}l_{i^{\prime }}n_{j^{\prime
   }}l_{j^{\prime }}\right) }
   \nonumber  \\[1ex]
   & &  =
   2k\left( k+1\right) \left( T^{k+1}\left(
   n_il_in_jl_j,n_{i^{\prime }}l_{i^{\prime }}n_{j^{\prime }}l_{j^{\prime
   }}\right) -T^{k-1}\left( n_il_in_jl_j,n_{i^{\prime }}l_{i^{\prime
   }}n_{j^{\prime }}l_{j^{\prime }}\right) \right)  
   \nonumber  \\[1ex]
   & &  +
   \left( l_i\left( l_i+1\right) -k\left( k+1\right) -l_{i^{\prime }}\left(
   l_{i^{\prime }}+1\right) \right) \left( U^{k+1}\left(
   n_il_in_jl_j,n_{i^{\prime }}l_{i^{\prime }}n_{j^{\prime }}l_{j^{\prime
   }}\right) -U^{k-1}\left( n_il_in_jl_j,n_{i^{\prime }}l_{i^{\prime
   }}n_{j^{\prime }}l_{j^{\prime }}\right) \right)
   \nonumber  \\[1ex]
   & &  +
   \left( l_j\left( l_j+1\right) -k\left( k+1\right) -l_{j^{\prime }}\left(
   l_{j^{\prime }}+1\right) \right) \left( U^{k+1}\left(
   n_jl_jn_il_i,n_{j^{\prime }}l_{j^{\prime }}n_{i^{\prime }}l_{i^{\prime 
   }}\right) - U^{k-1}\left( n_jl_jn_il_i,n_{j^{\prime }}l_{j^{\prime
   }}n_{i^{\prime }}l_{i^{\prime }}\right) \right)
   \nonumber  \\[1ex]
   & &  +
   \frac 12\left( l_i\left( l_i+1\right) -k\left( k+1\right) -l_{i^{\prime
   }}\left( l_{i^{\prime }}+1\right) \right) \left( l_j\left( l_j+1\right)
   -k\left( k+1\right) -l_{j^{\prime }}\left( l_{j^{\prime }}+1\right) \right)
   \nonumber  \\[1ex]
   & &  \times
   \left[
   \frac{k-2}{k\left( 2k-1\right) }\left( N^{k-2}\left(
   n_il_in_jl_j,n_{i^{\prime }}l_{i^{\prime }}n_{j^{\prime }}l_{j^{\prime
   }}\right) + N^{k-2}\left( n_jl_jn_il_i,n_{j^{\prime }}l_{j^{\prime
   }}n_{i^{\prime }}l_{i^{\prime }}\right) \right) \right.
   \nonumber \\[1ex]
   & &  \left. -
   \frac{k+3}{\left( k+1\right) \left( 2k+3\right) }\left( N^k\left(
   n_il_in_jl_j,n_{i^{\prime }}l_{i^{\prime }}n_{j^{\prime }}l_{j^{\prime
   }}\right) + N^k\left( n_jl_jn_il_i,n_{j^{\prime }}l_{j^{\prime }}n_{i^{\prime
   }}l_{i^{\prime }}\right) \right) \right].
\end{eqnarray}
The radial integrals are defined as
\begin{eqnarray}
\label{eq:Tensorial-oo-c}
\lefteqn{
   T^k\left( n_il_in_jl_j,n_{i^{\prime }}l_{i^{\prime }}n_{j^{\prime
   }}l_{j^{\prime }}\right) =\frac{\alpha ^2}{4\left( 2k+1\right) } }
   \nonumber \\[1ex]
   & & \times
   \int_0^\infty \int_0^\infty P_i\left( r_1\right) P_j\left(
   r_2\right) \frac{r_<^k}{r_>^{k+1}}
   \left(\frac{\partial}{\partial r_1}+\frac{1}{r_1}\right)
   P_{i^{\prime }}\left( r_1\right)
   \left(\frac{\partial}{\partial r_2}+\frac{1}{r_2}\right)
   P_{j^{\prime }}\left( r_2\right) dr_1dr_{2},
\end{eqnarray}
\begin{eqnarray}
\label{eq:Tensorial-oo-d}
\lefteqn{
   U^k\left( n_il_in_jl_j,n_{i^{\prime }}l_{i^{\prime }}n_{j^{\prime
   }}l_{j^{\prime }}\right) =\frac{\alpha ^2}{4\left( 2k+1\right) } }
   \nonumber \\[1ex]
   & & \times
   \int_0^\infty \int_0^\infty P_i\left( r_1\right) P_j\left(
   r_2\right)\left( 
   (k-1)\frac{r_2^k}{r_1^{k+2}}\epsilon (r_1-r_2)
   -(k+2)\frac{r_1^{k-1}}{r_2^{k+1}}\epsilon (r_2-r_1)\right)
   \nonumber \\[1ex]
   & & \times
   P_{i^{\prime }}\left( r_1\right)
   \left(\frac{\partial}{\partial r_2}+\frac{1}{r_2}\right)
   P_{j^{\prime }}\left( r_2\right) dr_1dr_{2} ,
\end{eqnarray}
\begin{eqnarray}
\label{eq:m-j}
\lefteqn{
   N^k\left( n_il_in_jl_j,n_{i^{\prime }}l_{i^{\prime }}n_{j^{\prime
   }}l_{j^{\prime }}\right) }
   \nonumber   \\[1ex]
   & & =
   \frac{\alpha ^2}4\int_0^\infty \int_0^\infty P_i\left( r_1\right) P_j\left(
   r_2\right) \frac{r_2^k}{r_1^{k+3}}\epsilon (r_1-r_2)P_{i^{\prime }}\left(
   r_1\right) P_{j^{\prime }}\left( r_2\right) dr_1dr_{2} ,
\end{eqnarray}
The integrals $N^k\left( n_il_in_jl_j,n_{i^{\prime }}l_{i^{\prime
}}n_{j^{\prime }}l_{j^{\prime }}\right) $, $T^k\left(
n_il_in_jl_j,n_{i^{\prime }}l_{i^{\prime }}n_{j^{\prime }}l_{j^{\prime
}}\right) $ and

$U^k\left( n_il_in_jl_j,n_{i^{\prime }}l_{i^{\prime }}n_{j^{\prime
}}l_{j^{\prime }}\right) $ have the following symmetry properties~\cite{Dan}:
\begin{eqnarray}
\label{eq:m-ka}
\lefteqn{
   N^k\left( n_il_in_jl_j,n_{i^{\prime }}l_{i^{\prime }}n_{j^{\prime
   }}l_{j^{\prime }}\right) }
   \nonumber \\
   &  & =
   N^k\left( n_{i^{\prime }}l_{i^{\prime
   }}n_{j^{\prime }}l_{j^{\prime }},n_il_in_jl_j\right)
   =N^k\left( n_{i^{\prime }}l_{i^{\prime }}n_jl_j,n_il_in_{j^{\prime
   }}l_{j^{\prime }}\right) 
   \nonumber \\
   &  & =
   N^k\left( n_il_in_{j^{\prime }}l_{j^{\prime
   }},n_{i^{\prime }}l_{i^{\prime }}n_jl_j\right) ,
\end{eqnarray}
\begin{eqnarray}
\label{eq:g_sym_t}
   T^k\left( n_il_in_jl_j,n_{i^{\prime }}l_{i^{\prime }}n_{j^{\prime }}l_{j^{\prime }}\right) 
   = T^k\left(n_{j}l_{j}n_{i}l_{i},n_{j^{\prime }}l_{j^{\prime }}n_{i^{\prime }}
   l_{i^{\prime }}\right) ,
\end{eqnarray}
\begin{eqnarray}
\label{eq:g_sym_u}
   U^k\left( n_il_in_jl_j,n_{i^{\prime }}l_{i^{\prime }}n_{j^{\prime }}l_{j^{\prime }}\right)
   =U^k\left( n_{i^{\prime }}l_{i^{\prime }}n_{j}l_{j},n_il_in_{j^{\prime }}l_{j^{\prime }}
   \right) .
\end{eqnarray}
As is seen from the above expressions, the symmetry of the $T^k$ and $U^k$ integrals is much
more restricted as compared to the $N^k$ integral.
There are some useful relations between these type of integrals, namely
\cite{Dan}:
\begin{eqnarray}
\label{eq:relat-TU}
  T^k\left( n_il_in_jl_j,n_{i^{\prime }}l_{i^{\prime }}n_{j^{\prime }}l_{j^{\prime }}\right) 
  +
  T^k\left( n_{i^{\prime }}l_{i^{\prime }}n_jl_j,n_il_in_{j^{\prime }}l_{j^{\prime }}\right) 
  =
  U^k\left( n_il_in_jl_j,n_{i^{\prime }}l_{i^{\prime }}n_{j^{\prime }}l_{j^{\prime }}\right) ,
\end{eqnarray}
\begin{eqnarray}
\label{eq:relat-UN}
\lefteqn{
   U^k \left( n_il_in_jl_j,n_{i^{\prime }}l_{i^{\prime }}n_{j^{\prime }}l_{j^{\prime }}\right)
   +
   U^k \left( n_{i^{\prime }}l_{i^{\prime }}n_{j^{\prime }}l_{j^{\prime }},n_il_in_jl_j\right)}
\nonumber  \\[1ex]
   & &   =
   - \frac{(k-1)(k+2)}{2k+1} \left\{
   N^{k-1}
   \left( n_il_in_jl_j,n_{i^{\prime }}l_{i^{\prime }}n_{j^{\prime }}l_{j^{\prime }}\right)
   +
   N^{k-1}
   \left( n_jl_jn_il_i,n_{j^{\prime }}l_{j^{\prime }}n_{i^{\prime }}l_{i^{\prime }}\right)
   \right\}
\nonumber  \\[1ex]
   & & +
   A \left( n_il_in_jl_j,n_{i^{\prime }}l_{i^{\prime }}n_{j^{\prime }}l_{j^{\prime }}\right),
\end{eqnarray}
where
\begin{eqnarray}
\label{eq:integ_A}
   A \left( n_il_in_jl_j,n_{i^{\prime }}l_{i^{\prime }}n_{j^{\prime }}l_{j^{\prime }}\right)
   =
   \frac{\alpha ^2}{4}\int_0^\infty R_i\left( r\right) R_j\left(r\right) 
   R_{i^{\prime }}\left( r\right) R_{j^{\prime }}\left( r\right) r^2 dr .
\end{eqnarray}
We will use these relations in section \ref{Simplification}
for getting the simplified expressions for submatrix elements of 
orbit--orbit operator.

\medskip

As is seen from the expressions (\ref{eq:Tensorial-oo-b}) and
(\ref{eq:i-a}), the matrix elements of $H_{oo1}^{(kk0,00)}$, $H_{oo2}^{(kk0,00)}$ 
and $H_{oo4}^{(kk0,00)}$ have the same angular dependence as the 
electrostatic (Coulomb) electron interaction operator $H^{Coulomb}$.
So it is most convenient to evaluate these
three terms simultaneously with the
electrostatic electron interaction operator \cite{E} which
itself contains the same tensorial structure 
\begin{eqnarray}
\label{eq:Tensorial-co-a}
   {\cal H}^{Coulomb} \equiv
\displaystyle {\sum_{k}} H_{Coulomb}^{(kk0,000)}
\end{eqnarray}
and its submatrix element is
\begin{eqnarray}
\label{eq:Tensorial-co-b}
\lefteqn{
  \left( n_i\lambda _in_j\lambda _j\left\| H_{Coulomb}^{(kk0,000)}
   \right\| n_{i^{\prime }}\lambda _{i^{\prime }}n_{j^{\prime }}\lambda
   _{j^{\prime }}\right) }
   \nonumber  \\[1ex]
   & &  = 
   2 [k]^{1/2}
   \left( l_i\left\| C^{\left( k \right) }\right\| l_{i^{\prime }}\right)
   \left( l_j\left\| C^{\left( k \right) }\right\| l_{j^{\prime }}\right)
   R_{k}\left( n_il_i n_{i^{\prime }}l_{i^{\prime }},n_jl_j n_{j^{\prime
   }}l_{j^{\prime }}\right).
\end{eqnarray}
So submatrix element for $H_{Coulomb}^{(kk0,000)}$, $H_{oo1}^{(kk0,000)}$,
$H_{oo2}^{(kk0,000)}$ and $H_{oo4}^{(kk0,000)}$ is
\begin{eqnarray}
\label{eq:Tensorial-coo-a}
\lefteqn{
   \left( n_i\lambda _in_j\lambda _j\left\|
   H_{Coulomb}^{(kk0,000)} + H_{oo1}^{(kk0,000)} + H_{oo2}^{(kk0,000)} +
   H_{oo4}^{(kk0,000)}
   \right\| n_{i^{\prime }}\lambda _{i^{\prime
   }}n_{j^{\prime }}\lambda _{j^{\prime }}\right) }
   \nonumber \\[1ex]
   & & =
   2 [k]^{1/2}
   \left( l_i\left\| C^{\left( k \right) }\right\| l_{i^{\prime }}\right)
   \left( l_j\left\| C^{\left( k \right) }\right\| l_{j^{\prime }}\right)
   \nonumber \\[1ex]
   & & \times
   \left\{ R_k\left(
   n_il_in_jl_j,n_{i^{\prime }}l_{i^{\prime }}n_{j^{\prime }}l_{j^{\prime
   }}\right) 
   -\left( 1-\delta \left( k,0\right) \right) ~Z_k\left(
   n_il_in_jl_j,n_{i^{\prime }}l_{i^{\prime }}n_{j^{\prime }}l_{j^{\prime
   }}\right) \right\} .
\end{eqnarray}
It is more convenient the remaining term $H_{oo3}^{(kk0,00)}$ to
calculate separately. Its matrix element is
\begin{eqnarray}
\label{eq:mg-a}
\lefteqn{
   \left( n_il_in_jl_j||H_{oo3}^{(kk0,00)}||n_{i^{\prime }}l_{i^{\prime}}n_{j^{\prime 
   }}l_{j^{\prime }}\right) }
   \nonumber  \\[1ex]
   & & =
   2 \sqrt{2k+1}\frac 1{k(k+1)} 
   \left( \left( l_i+l_{i^{\prime}}+k+2\right) \left( l_i+l_{i^{\prime }}-k\right) 
   \left( l_i-l_{i^{\prime }}+k+1\right) \right. 
   \nonumber  \\[1ex]
   & & \times
   \left. \left( l_{i^{\prime }}-l_i+k+1\right)
   \left( l_j+l_{j^{\prime }}+k+2\right)
   \times \left( l_j+l_{j^{\prime }}-k\right) \left( l_j-l_{j^{\prime
   }}+k+1\right)  \right. 
   \nonumber  \\[1ex]
   & & \times
   \left. \left( l_{j^{\prime }}-l_j+k+1\right) \right) ^{1/2}
   \left( l_i||C^{\left( k+1\right) }||l_{i^{\prime }}\right) \left(
   l_j||C^{\left( k+1\right) }||l_{j^{\prime }}\right)
   \nonumber  \\[1ex]
   & & \times
   \left( N^{k-1}\left(
   n_il_in_jl_j,n_{i^{\prime }}l_{i^{\prime }}n_{j^{\prime }}l_{j^{\prime
   }}\right) +N^{k-1}\left( n_jl_jn_il_i,n_{j^{\prime }}l_{j^{\prime
   }}n_{i^{\prime }}l_{i^{\prime }}\right) \right) .
\end{eqnarray}
The use of the approach presented in \cite{Gaigalas-meth2}
presumes that both the tensorial structure of the operator
under consideration and the submatrix elements $\left( n_i\lambda
_in_j\lambda _j\left\| g\right\| n_{i^{\prime }}\lambda _{i^{\prime
}}n_{j^{\prime }}\lambda _{j^{\prime }}\right) $ are known. 
The formulae (\ref{eq:Tensorial-oo-b}) or (\ref{eq:Tensorial-coo-a}) and 
(\ref{eq:mg-a}) are the 
expressions we need. We may readily obtain the value
of a matrix element of this operator for any number of open shells in bra
and ket functions, by choosing tensorial structure from 
(\ref{eq:b}), using their submatrix elements in an expression of the type 
(\ref{eq:Tensorial-d}), defining bra and ket functions, and
performing spin--angular integrations according to \cite{Gaigalas-meth2}.

\section{Some simplification for submatrix elements}
\label{Simplification}

In this section we will discuss some special cases of distributions 
$i j i^{\prime } j^{\prime }$ from Gaigalas {\it et al.}~\cite{Gaigalas-meth2}
for the orbit--orbit interaction operator.

\medskip

Let us at first consider the distribution
$i j i^{\prime } j^{\prime }= \alpha \beta \alpha \beta$.
Using the (\ref{eq:i-a}) we express the coefficient $Z_k$ as 
\begin{eqnarray}
\label{eq:Z-abab-a}
   Z_k\left( n_{\alpha}l_{\alpha}n_{\beta}l_{\beta},
   n_{\alpha}l_{\alpha}n_{\beta}l_{\beta}\right) 
   =
   Z_k^{\prime}\left( n_{\alpha}l_{\alpha}n_{\beta}l_{\beta},
   n_{\alpha}l_{\alpha}n_{\beta}l_{\beta}\right) 
   +
   Z_k^{\prime \prime}\left( n_{\alpha}l_{\alpha}n_{\beta}l_{\beta},
   n_{\alpha}l_{\alpha}n_{\beta}l_{\beta}\right) ,
\end{eqnarray}
where
\begin{eqnarray}
\label{eq:Z-abab-b}
\lefteqn{
   Z_k^{\prime}\left( n_{\alpha}l_{\alpha}n_{\beta}l_{\beta},
   n_{\alpha}l_{\alpha}n_{\beta}l_{\beta}\right)} 
   \nonumber \\[1ex]
   & &  =
   k\left( k+1\right) 
   \left[  \: 2 T^{k+1}\left( n_{\alpha}l_{\alpha}n_{\beta}l_{\beta},
   n_{\alpha}l_{\alpha}n_{\beta}l_{\beta}\right)   \right.
   \nonumber \\[1ex]
   & & -
   U^{k+1}\left( n_{\alpha}l_{\alpha}n_{\beta}l_{\beta},
   n_{\alpha}l_{\alpha}n_{\beta}l_{\beta}\right)
   - U^{k+1}\left( n_{\beta}l_{\beta}n_{\alpha}l_{\alpha},
   n_{\beta}l_{\beta}n_{\alpha}l_{\alpha} \right)
   \nonumber \\[1ex]
   & & -
   \left.
   \frac{k\left( k+1\right)\left( k+3\right)}{\left( k+1\right) \left( 2k+3\right) }
   \left(
   N^k\left( n_{\alpha}l_{\alpha}n_{\beta}l_{\beta},
   n_{\alpha}l_{\alpha}n_{\beta}l_{\beta}\right)
   + N^k\left( n_{\beta}l_{\beta}n_{\alpha}l_{\alpha}
   n_{\beta}l_{\beta}n_{\alpha}l_{\alpha}\right)
   \right) \right] ,
\end{eqnarray}
\begin{eqnarray}
\label{eq:Z-abab-c}
\lefteqn{
   Z_k^{\prime \prime}\left( n_{\alpha}l_{\alpha}n_{\beta}l_{\beta},
   n_{\alpha}l_{\alpha}n_{\beta}l_{\beta}\right)} 
   \nonumber \\[1ex]
   & &  =
   - k\left( k+1\right) 
   \left[ \: 2 T^{k-1}\left( n_{\alpha}l_{\alpha}n_{\beta}l_{\beta},
   n_{\alpha}l_{\alpha}n_{\beta}l_{\beta}\right) \right.
   \nonumber \\[1ex]
   & & -
   U^{k-1} \left( n_{\alpha}l_{\alpha}n_{\beta}l_{\beta},
   n_{\alpha}l_{\alpha}n_{\beta}l_{\beta}\right) 
   - U^{k-1}\left( n_{\beta}l_{\beta}n_{\alpha}l_{\alpha},
   n_{\beta}l_{\beta}n_{\alpha}l_{\alpha}\right) 
   - 
   \frac{k\left( k+1\right)\left(k-2\right)}{k\left(2k-1\right)}
    \nonumber \\[1ex]
   & & \times
   \left.   \left(
   N^{k-2}\left( n_{\alpha}l_{\alpha}n_{\beta}l_{\beta},
   n_{\alpha}l_{\alpha}n_{\beta}l_{\beta}\right)
   + N^{k-2}\left( n_{\beta}l_{\beta}n_{\alpha}l_{\alpha},
   n_{\beta}l_{\beta}n_{\alpha}l_{\alpha}\right)
   \right) \right] .
\end{eqnarray}
Let us start to evaluate the expression (\ref{eq:Z-abab-b}). 
We can rewrite the
$T^{k+1}\left( n_{\alpha}l_{\alpha}n_{\beta}l_{\beta},
   n_{\alpha}l_{\alpha}n_{\beta}l_{\beta}\right)$ using the (\ref{eq:relat-TU})
as
\begin{eqnarray}
\label{eq:Z-abab-d}
\lefteqn{
   2 T^{k+1}\left( n_{\alpha}l_{\alpha}n_{\beta}l_{\beta},
   n_{\alpha}l_{\alpha}n_{\beta}l_{\beta}\right) }
   \nonumber \\[1ex]
   & & =
   \left[ T^{k+1}\left( n_{\alpha}l_{\alpha}n_{\beta}l_{\beta},
   n_{\alpha}l_{\alpha}n_{\beta}l_{\beta}\right)
   +    T^{k+1}\left( n_{\alpha}l_{\alpha}n_{\beta}l_{\beta},
   n_{\alpha}l_{\alpha}n_{\beta}l_{\beta}\right) \right]
   \nonumber \\[1ex]
   & & =
   U^{k+1}\left( n_{\alpha}l_{\alpha}n_{\beta}l_{\beta},
   n_{\alpha}l_{\alpha}n_{\beta}l_{\beta}\right) .
\end{eqnarray}
With the help of equation (\ref{eq:relat-UN}) we are rewriting the 
$U^{k+1}\left( n_{\beta}l_{\beta}n_{\alpha}l_{\alpha},
   n_{\beta}l_{\beta}n_{\alpha}l_{\alpha}\right) $ as
\begin{eqnarray}
\label{eq:Z-abab-e}
\lefteqn{
   U^{k+1}\left( n_{\beta}l_{\beta}n_{\alpha}l_{\alpha},
   n_{\beta}l_{\beta}n_{\alpha}l_{\alpha}\right) }
   \nonumber \\[1ex]
   & & =
   \frac{1}{2} \left[ U^{k+1}\left( n_{\beta}l_{\beta}n_{\alpha}l_{\alpha},
   n_{\beta}l_{\beta}n_{\alpha}l_{\alpha}\right)
   + U^{k+1}\left( n_{\beta}l_{\beta}n_{\alpha}l_{\alpha},
   n_{\beta}l_{\beta}n_{\alpha}l_{\alpha}\right) \right]
   \nonumber \\[1ex]
   & & = -
   \frac{k\left(k+3\right)}{2k+3}
   \left[ N^{k}\left( n_{\beta}l_{\beta}n_{\alpha}l_{\alpha},
   n_{\beta}l_{\beta}n_{\alpha}l_{\alpha}\right)
   + N^{k}\left( n_{\beta}l_{\beta}n_{\alpha}l_{\alpha},
   n_{\beta}l_{\beta}n_{\alpha}l_{\alpha}\right) \right]
   \nonumber \\[1ex]
   & &  +
   A \left( n_{\beta}l_{\beta}n_{\alpha}l_{\alpha},
   n_{\beta}l_{\beta}n_{\alpha}l_{\alpha}\right) .
\end{eqnarray}
So inserting equations (\ref{eq:Z-abab-d})
and (\ref{eq:Z-abab-e}) in the (\ref{eq:Z-abab-b}) we have:
\begin{eqnarray}
\label{eq:Z-abab-f}
   Z_k^{\prime}\left( n_{\alpha}l_{\alpha}n_{\beta}l_{\beta},
   n_{\alpha}l_{\alpha}n_{\beta}l_{\beta}\right) =
   - \frac{k\left( k+1 \right)}{2}
   A \left( n_{\beta}l_{\beta}n_{\alpha}l_{\alpha},
   n_{\beta}l_{\beta}n_{\alpha}l_{\alpha}\right) .
\end{eqnarray}
After similar rearrangements of the expression (\ref{eq:Z-abab-c}) we have:
\begin{eqnarray}
\label{eq:Z-abab-g}
   Z_k^{\prime \prime}\left( n_{\alpha}l_{\alpha}n_{\beta}l_{\beta},
   n_{\alpha}l_{\alpha}n_{\beta}l_{\beta}\right) =
   \frac{k\left( k+1 \right)}{2}
   A \left( n_{\beta}l_{\beta}n_{\alpha}l_{\alpha},
   n_{\beta}l_{\beta}n_{\alpha}l_{\alpha}\right) .
\end{eqnarray}
So finally
\begin{eqnarray}
\label{eq:Z-abab-h}
   Z_k\left( n_{\alpha}l_{\alpha}n_{\beta}l_{\beta},
   n_{\alpha}l_{\alpha}n_{\beta}l_{\beta}\right) = 0
\end{eqnarray}
or
\begin{eqnarray}
\label{eq:Tensorial-oo-abab}
   \left( n_{\alpha}\lambda _{\alpha}n_{\beta}\lambda _{\beta}\left\|
   H_{oo1}^{(kk0,000)} + H_{oo2}^{(kk0,000)} +
   H_{oo4}^{(kk0,000)}
   \right\| n_{\alpha}\lambda _{\alpha}n_{\beta}\lambda _{\beta}\right) = 0 .
\end{eqnarray}
It means that for distributions 
$i j i^{\prime } j^{\prime }= \alpha \beta \alpha \beta$ we do
not need to calculate matrix elements of the terms
$H_{oo1}^{(kk0,000)}$, $H_{oo2}^{(kk0,000)}$ and $H_{oo4}^{(kk0,000)}$ at all.
In a similar way it is possible to prove that
\begin{eqnarray}
\label{eq:Z-all_dis}
\lefteqn{
   Z_k\left( n_{\alpha}l_{\alpha}n_{\alpha}l_{\alpha},
   n_{\alpha}l_{\alpha}n_{\alpha}l_{\alpha}\right) = 
   Z_k\left( n_{\beta}l_{\beta}n_{\alpha}l_{\alpha},
   n_{\beta}l_{\beta}n_{\alpha}l_{\alpha}\right) = 
   Z_k\left( n_{\beta}l_{\beta}n_{\alpha}l_{\alpha},
   n_{\alpha}l_{\alpha}n_{\alpha}l_{\alpha}\right) }
   \nonumber \\[1ex]
   &  & =
   Z_k\left( n_{\alpha}l_{\alpha}n_{\beta}l_{\beta},
   n_{\alpha}l_{\alpha}n_{\alpha}l_{\alpha}\right) = 
   Z_k\left( n_{\beta}l_{\beta}n_{\beta}l_{\beta},
   n_{\alpha}l_{\alpha}n_{\beta}l_{\beta}\right) = 
   Z_k\left( n_{\beta}l_{\beta}n_{\beta}l_{\beta},
   n_{\alpha}l_{\alpha}n_{\beta}l_{\beta}\right) 
   \nonumber \\[1ex]
   &  & =
   Z_k\left( n_{\beta}l_{\beta}n_{\gamma}l_{\gamma},
   n_{\alpha}l_{\alpha}n_{\gamma}l_{\gamma}\right) = 
   Z_k\left( n_{\beta}l_{\beta}n_{\gamma}l_{\gamma},
   n_{\alpha}l_{\alpha}n_{\gamma}l_{\gamma}\right) = 0 .
\end{eqnarray}
So for the distributions
$\alpha \alpha \alpha \alpha$, $\alpha \beta \alpha \beta$,
$\beta \alpha \beta \alpha$,
$\beta \alpha \alpha \alpha$, $\alpha \beta \alpha \alpha$,
$\beta \beta \beta \alpha$, $\beta \beta \alpha \beta$,
$\beta \gamma \alpha \gamma$, $\gamma \beta \gamma \alpha$ we do
not need to calculate matrix elements of $H_{oo1}^{(kk0,000)}$, 
$H_{oo2}^{(kk0,000)}$ and $H_{oo4}^{(kk0,000)}$ terms, too.
In these cases the orbit--orbit interaction operator contains the
term $H_{oo3}^{(kk0,000)}$ only. 
The matrix element of this term has the radial integral of only one type,
i.e. $N^{k-1} \left( n_jl_jn_il_i,n_{j^{\prime }}l_{j^{\prime }}n_{i^{\prime 
}}l_{i^{\prime }}\right)$. 

\medskip

It is very well known in the literature \cite{JS} that the matrix elements of
the orbit-orbit operator ${\cal H}_{OO}$,
$(ns^2\;^1S||{\cal H}_{OO}||ns^2\;^1S)$ and
$(ns\;n^{\prime }s\;^1S||{\cal H}_{OO}||ns\;n^{\prime }s\;^1S)$,
are zeroes. It is possible to generalize these statements using the results 
of present paper.
We see that for direct part of any diagonal
matrix elements or the off--diagonal matrix of the type

$(...nl^{N}...n^{\prime }l^{\prime N^{\prime }}...\;LS||{\cal H}_{OO}||
...nl^{N \pm 1}...n^{\prime }l^{\prime N^{\prime }\mp 1}...\;L^{\prime 
}S^{\prime })$ we need to calculate the matrix element of 
$H_{oo3}^{(kk0,000)}$ operator only.
Using the fact that $\left( 0 || C^{(1)} || 0 \right) = 0$, we strightforwardly
from (\ref{eq:mg-a}) find values of these matrix elements 
in the case $l,l^{\prime}$ = 0. These values and values of exchange part
of diagonal matrix elements are equal to zero in this case.
%But if $l,l^{\prime}$ = 0 then matrix element of orbit--orbit interaction 
%operator equal to zero. 
This is valid for matrix elements between functions with any number of open 
electron shells.

\medskip

Remaining 33 distributions from Table~1 of
Gaigalas {\it et al.}~\cite{Gaigalas-meth2} have all terms
$H_{oo1}^{(kk0,000)}$, $H_{oo2}^{(kk0,000)}$, $H_{oo3}^{(kk0,000)}$ and 
$H_{oo4}^{(kk0,000)}$. For calculation of matrix elements of these distributions
we need to find the values of 
$T^{k\pm1} \left( n_jl_jn_il_i,n_{j^{\prime }}l_{j^{\prime }}n_{i^{\prime 
}}l_{i^{\prime }}\right)$,
$U^{k\pm1} \left( n_jl_jn_il_i,n_{j^{\prime }}l_{j^{\prime }}n_{i^{\prime 
}}l_{i^{\prime }}\right)$,

$N^{k-1} \left( n_jl_jn_il_i,n_{j^{\prime }}l_{j^{\prime }}n_{i^{\prime 
}}l_{i^{\prime }}\right)$,
$N^{k-2} \left( n_jl_jn_il_i,n_{j^{\prime }}l_{j^{\prime }}n_{i^{\prime 
}}l_{i^{\prime }}\right)$
and
$N^{k} \left( n_jl_jn_il_i,n_{j^{\prime }}l_{j^{\prime }}n_{i^{\prime 
}}l_{i^{\prime }}\right)$
integrals (see (\ref{eq:i-a}) and (\ref{eq:mg-a})).

\section{The effect of the orbit--orbit interaction on ground states
in light atoms}

Taking into account the relativistic corrections in the
Breit--Pauli approximation in the configuration interaction method (CI),
it is important to know the matrix elements of operators considered.
As was shown in the Section 4 the matrix elements of
the orbit--orbit operator ${\cal H}_{OO}$,
$(ns^2\;^1S||{\cal H}_{OO}||ns^2\;^1S)$ and
$(ns\;n^{\prime }s\;^1S||{\cal H}_{OO}||ns\;n^{\prime }s\;^1S)$,
are zeroes. Therefore, in this approximation for the configuration
$1s^2\;^1S$ the corrections due to orbit--orbit operator appear
through the diagonal matrix elements of the types
$(np^2\;^1S||{\cal H}_{OO}||np^2\;^1S)$,
$(nd^2\;^1S||{\cal H}_{OO}||nd^2\;^1S)$, etc., and through the off--diagonal
matrix elements. In investigating the level
$1s^2ns\;^2S$, the matrix elements
$(ns^2n^{\prime }s\;^2S||{\cal H}_{OO}||ns^2n^{\prime }s\;^2S)$,
of the orbit--orbit operator are equal to zero, too. So the orbit--orbit operator
corrections appear through the remaining matrix elements where the resulting
terms of bra and ket functions coincide. Therefore it is plausible that these
corrections are unimportant to the absolute values
of the level $1s^2\;^1S$.
Of course, one has to investigate into their exact contribution, as
compared to other relativistic corrections, in aiming at high accuracy of
the results.

\subsection{The MCHF method with Breit--Pauli and mass--polarization
corrections}

The computational method for including nuclear and relativistic effects has
been described in detail elsewhere~\cite{book}. Briefly, the wave function 
$\Psi(\gamma LS)$ for an atomic state labelled by the configuration 
$\gamma$,
and term $LS$ is approximated by a linear combination of configuration state
functions (CSFs),
\begin{equation}
\Psi(\gamma LS) = \sum_{i=1}^{M} c_{i}\Phi(\gamma_i LS).
\end{equation}
Each $\Phi(\gamma_i LS)$ is constructed from one--electron spin--orbitals for
the configuration $\gamma_i$ and is of the same $LS$ symmetry as the atomic
state function. In the MCHF method, the radial functions used to construct
the CSFs and the expansion coefficients $c_i$ are determined variationally
so as to leave the non--relativistic energy stationary with respect to
variations in the radial functions and the expansion 
coefficients~\cite{book}. Once radial functions have been determined, 
they may be used as a basis
for a wave function expansion including additional effects. In particular,
when relativistic corrections are included,
\begin{equation}
\Psi(\gamma LSJ) = \sum_{LS}\sum_{i=1}^{M_{LS}} c_{i,LS}\Phi(\gamma_i LSJ).
\end{equation}
where $\Phi(\gamma LSJ)$ is a CSF in which the total angular momentum $L$
and the total spin $S$ have been coupled to a resultant $J$. The expansion
coefficients are obtained from a configuration interaction calculation,
where the interaction matrix is evaluated with respect to the Breit--Pauli
Hamiltonian, or some subset of operators. 
New, efficient programs based on the combination
of second quantization in coupled tensorial form, and a generalized
graphical technique~\cite{Gaigalas-meth2} were used for evaluating the 
Breit--Pauli operators. The expressions for orbit--orbit interaction operator
are taken from present work.

\subsection{Ground states and ionization potentials for Li--like atoms
and ions}

\begin{table}
\caption{Comparison of contributions to the Breit--Pauli
energies (in au)  between MCHF (present work) and the full core
plus correlation results of Chung~\cite{C1}
(the second line).}
\begin{center}
\begin{tabular}{l r r r }
\hline \hline
& $1s^22s\;^2\!S$ & $1s^2\;^1S$ & IP \\
\hline
\multicolumn{3}{c}{\sl Li~I} & \\
$E_{nr}{}^a $           & -7.4779329 & -7.2798008 & 0.1981322 \\
                        & -7.4779251 & -7.2797824 & 0.1981579\\
$E_{RS-oo}{}^a $        & -0.0005924 & -0.0005811 & 0.0000113\\
                        & -0.0005886 & -0.0005773 & 0.0000111\\
$E_{oo}{}^a $           & -0.0000234 & -0.0000230 & 0.0000004\\
                        & -0.0000233 & -0.0000229 & 0.0000004\\
$E_{QED}$ \cite{C2}     &            &            &-0.0000004\\
\multicolumn{3}{c}{\sl Be~II} & \\
$E_{nr}{}^a $           &-14.3246101 &-13.6554354 & 0.6691747\\
                        &-14.3246043 &-13.6554171 & 0.6691872\\
$E_{RS-oo}{}^a $        & -0.0022620 & -0.0021650 & 0.0000970\\
                        & -0.0022362 & -0.0021404 & 0.0000958\\
$E_{oo}{}^a $           & -0.0000485 & -0.0000468 & 0.0000017\\
                        & -0.0000486 & -0.0000470 & 0.0000017\\
$E_{QED}$ \cite{C2}     &            &            &-0.0000035\\
\multicolumn{3}{c}{\sl B~III} & \\
$E_{nr}{}^a $           &-23.4244364 &-22.0308301 & 1.3936063\\
                        &-23.4244328 &-22.0308116 & 1.3936211\\
$E_{RS-oo}{}^a $        & -0.0062014 & -0.0058314 & 0.0003700\\
                        & -0.0060953 & -0.0057303 & 0.0003648\\
$E_{oo}{}^a $           & -0.0000835 & -0.0000798 & 0.0000037\\
                        & -0.0000834 & -0.0000796 & 0.0000037\\
$E_{QED}$ \cite{C2}     &            &            &-0.0000135\\
\multicolumn{3}{c}{\sl C~IV} & \\
$E_{nr}{}^a $           &-34.7753307 &-32.4060978 & 2.3692330\\
                        &-34.7753254 &-32.4060767 & 2.3692487\\
$E_{RS-oo}{}^a $        & -0.0139422 & -0.0129474 & 0.0009948\\
                        & -0.0136083 & -0.0126276 & 0.0009808\\
$E_{oo}{}^a $           & -0.0001286 & -0.0001218 & 0.0000067\\
                        & -0.0001275 & -0.0001209 & 0.0000066\\
$E_{QED}$ \cite{C2}     &            &            &-0.0000350\\
\hline \\ [-0.3cm]
   $ {}^a $ This work. & & & \\
\end{tabular}
\end{center}
\end{table}

\begin{table}  
Table~1: (continued)
%Comparison of contributions to the Breit-Pauli
%energies (in au)  between MCHF (present work) and the full core
%plus correlation results of Chung~\cite{C1} are reported on
%the second line.
\vspace{5mm}\\
\begin{center}
\begin{tabular}{l r r r }
\hline \hline
& $1s^22s\;^2\!S$ & $1s^2\;^1S$ & IP \\
\hline
\multicolumn{3}{c}{\sl N~V} & \\
$E_{nr}{}^a $           &-48.3767096 &-44.7812909 & 3.5954187\\
                        &-48.3767060 &-44.7812707 & 3.5954353\\
$E_{RS-oo}{}^a $        & -0.0274761 & -0.0252846 & 0.0021915\\
                        & -0.0265949 & -0.0244425 & 0.0021523\\
$E_{oo}{}^a $           & -0.0001831 & -0.0001726 & 0.0000105\\
                        & -0.0001812 & -0.0001708 & 0.0000103\\
$E_{QED}$ \cite{C2}     &            &            &-0.0000734\\
\multicolumn{3}{c}{\sl O~VI} & \\
$E_{nr}{}^a $           &-64.2283470 &-59.1564366 & 5.0719104\\
                        &-64.2283436 &-59.1564162 & 5.0719275\\
$E_{RS-oo}{}^a $        & -0.0492822 & -0.0450456 & 0.0042366\\
                        & -0.0472561 & -0.0430947 & 0.0041412\\
$E_{oo}{}^a $           & -0.0002473 & -0.0002322 &  0.0000151\\
                        & -0.0002442 & -0.0002294 &  0.0000148\\
$E_{QED}$ \cite{C2}     &            &            & -0.0001344 \\
\multicolumn{3}{c}{\sl F~VII} & \\
$E_{nr}{}^a $           &-82.3301381 &-75.5315505 & 6.7985876\\
                        &-82.3301340 &-75.5315288 & 6.7986052\\
$E_{RS-oo}{}^a $        & -0.0824020 & -0.0749334 & 0.0074686\\
                        & -0.0780867 & -0.0708284 & 0.0072583\\
$E_{oo}{}^a $           & -0.0003222 & -0.0003015 & 0.0000207\\
                        & -0.0003168 & -0.0002966 & 0.0000202\\
$E_{QED}$ \cite{C2}     &            &            &-0.0002240\\
\hline \\ [-0.3cm]
   $ {}^a $ This work. & & & \\

\end{tabular}
\end{center}
\end{table}

The ground states of Li isoelectronic sequency was found using method 
described above. 
The calculations reported here are strictly {\sl ab initio}: no 
$l$--extrapolation or basis extrapolation has been applied. 
The configuration states included in the expansions of
different terms were obtained by including all possible CSFs of a given 
$LS$ symmetry that could be constructed from orbitals with $n < 10$, 
$l < 7$.
The largest expansion for this rule--based scheme was for
$1s^22s\;^2\!S$, where the interaction matrix size was 7~496.

\medskip

In Table 1, we compare contributions to the Breit--Pauli energies (without
the mass--polarization correction) of $1s^22s$ \Term 2 S {} / and $1s^2$
\Term
1 S {} / with those reported by Chung~\cite{C1}. We see that the
non-relativistic energies are in close agreement though, the present
results, without any extrapolations are slightly lower than those of Chung.
The non--relativistic ionization potential (IP) is in close agreement. The
relativistic shift effect is in surprisingly large disagreement
for B~III -- F~VII though,
again, the contribution to IP agrees more closely. Finally, the much smaller
orbit--orbit effect is in good agreement, with the difference again, agreeing
to more decimal places than the individual energies. Some differences are
expected since, in the present work, effects are included in the interaction
matrix, prior to matrix diagonalization whereas in the full--core plus
correlation method employed by Chung, these result are computed as a
small perturbative correction from the non--relativistic wave function.
However, in the present methodology, there also may be basis effects, in
that the orbitals used in the expansion are optimized for the
non--relativistic Hamiltonian and are incomplete with respect to the
Breit--Pauli Hamiltonian. This is particularly true for the relativistic
shift without orbit-orbit interaction where the one-electron Darwin term
depends only on the value of the $s$--orbitals at the nucleus. The effect of
orbit-orbit interaction on the energies is small, and the contribution to the ionization
potential is in good agreement between the two theories. Finally, to gain some
perspective on the magnitude of corrections, we include the QED correction
of $-0.0001344$ au. to the ionization potential of O~VI
as reported by Chung~\cite
{C2}. However, it should be noted that the orbit-orbit correction to the $2p$
ionization potential is -0.0003328 au. \cite{wang} so the relative
importance depends on the state. For a correct spectrum, it appears that
both should be included, at least for the lower levels.

\begin{table}
\caption{Comparison of $E_{nr}$, $E_{RS-oo}$, $E_{oo}$ energies 
for F~VII and F~VIII ground states in seven different expansions.
}
\begin{center}
\begin{tabular}{l r r r r r r r}
\hline \hline
 & \multicolumn{3}{c}{$1s^22s\;^2\!S$} & \quad &
 \multicolumn{3}{c}{$1s^2\;^1S$}\\
\cline{2-4}\cline{6-8}
CSF  & $E_{nr}$ & $E_{RS-oo}$ & $E_{oo}$ & & $E_{nr}$ & $E_{RS-oo}$ & $E_{oo}$ \\
\hline
$nl \le 3d$ & -82.3229146 & -.0786149 & -.0002951 & & -75.5261806 & -.0713471 & -.0002806\\
$nl \le 4f$ & -82.3275504 & -.0793346 & -.0003151 & & -75.5297942 & -.0719734 & -.0002958\\
$nl \le 5g$ & -82.3290016 & -.0799856 & -.0003195 & & -75.5306952 & -.0726407 & -.0002992\\
$nl \le 6h$ & -82.3296043 & -.0805299 & -.0003211 & & -75.5311425 & -.0731232 & -.0003006\\
$nl \le 7i$ & -82.3298946 & -.0811112 & -.0003218 & & -75.5313610 & -.0737041 & -.0003012\\
$nl \le 8k$ & -82.3300484 & -.0817376 & -.0003221 & & -75.5314792 & -.0742891 & -.0003014\\
$nl \le 9l$ & -82.3301381 & -.0824020 & -.0003222 & & -75.5315505 & -.0749334 & -.0003015\\
\hline
\end{tabular}
\end{center}
\end{table}

\begin{table}
\caption{Comparison of ionization potential (in au) for the $1s^22s\;^2\!S$ states of
F~VII in different expensions with different corrections.
$IP_{nr}$ -- nonrelativistic case,
$IP_{RS-oo}$ -- only relativistic shift without orbit--orbit term,
$IP_{oo}$ -- only orbit--orbit term.
}
\begin{center}
\begin{tabular}{l r r r r r r}
\hline \hline
 & \multicolumn{2}{c}{Number of CSF for } & \quad &
 & &\\
\cline{2-3}
CSF  & $1s^22s\;^2\!S$ & $1s^2\;^1S$ & & $IP_{nr}$ & $IP_{RS-oo}$ & $IP_{oo}$ \\
\hline
$nl \le 3d$ & 27   & 10  & & 6.7967340 & .0072678 & .0000145\\
$nl \le 4f$ & 110  & 20  & & 6.7977562 & .0073612 & .0000193\\
$nl \le 5g$ & 338  & 35  & & 6.7983064 & .0073449 & .0000203\\
$nl \le 6h$ & 866  & 56  & & 6.7984618 & .0074067 & .0000205\\
$nl \le 7i$ & 1948 & 84  & & 6.7985336 & .0074071 & .0000206\\
$nl \le 8k$ & 3974 & 120 & & 6.7985692 & .0074485 & .0000207\\
$nl \le 9l$ & 7496 & 165 & & 6.7985876 & .0074686 & .0000207\\
\hline
\end{tabular}
\end{center}
\end{table}

In Table~2 and Table~3, we report the contributions to the Breit--Pauli ground 
energies of F~VII and F~VIII and ionization potential
using seven different expansions (first column of Table~2 and Table~3). The notation
$nl \le 3d$, for example, means that the expansion was obtained by including all possible
CSFs of a given $LS$ symmetry that could be constructed from orbitals with $n \le 3$, $l \le 2$.
It contains 27 configurations for $1s^22s\;^2\!S$ and 10 configurations for $1s^2\;^1S$ 
(see Table 3).

\medskip

The Table~2 indicates that in case the CSF number is being increased,
non--relativistic energy $E_{nr}$ lowers.
The energy is also being lowered by $E_{RS-oo}$ and $E_{oo}$ corrections. Their absolute
values increase insignificantly at the increase of CSF number. Having
compared those results with Chung results, we notice that the values of $E_{RS-oo}$ and 
$E_{oo}$ indicated in the article tally best with the values of Chung, when
CSF are generated from orbital $nl \le 3d$ or $nl \le 4f$. Whereas 
non--relativistic energy values, got in the article, tally best, when CSF basis
includes orbitals $nl \le 9l$. The same discussions valid for ionization potentials in
different approximations (see Table~3), too.

\section{Conclusions}

The general irreducible tensorial form of the orbit--orbit interaction operator 
in the formalism of second quantization is presented
(expressions (\ref{eq:Tensorial-d}), (\ref{eq:Tensorial-oo-b})
(\ref{eq:mg-a}) ). It contains four different terms. Each is
associated with different set of radial integrals. In the present work
we have succeeded in obtaining simpler expressions having only the term
$H_{oo3}^{(kk0,000)}$, for some special distributions of electrons in the 
configuration. 
As we see from the paper the formalism for evaluation of matrix elements
developed by Gaigalas {\it et al.}~\cite{Gaigalas-meth2}
allow us to use these simplifications for practical applications in general way.
This facilitates practical
calculations of matrix elements without restraining the generality,
and is one more advantage of the approach used.
The properties (\ref{eq:Z-abab-h}) and (\ref{eq:Z-all_dis}) are useful 
for testing the calculation of matrix elements and for evaluation of the
accuracy of radial integrals $T^{k}$, $U^{k}$ and $N^{k}$, too.

\medskip

The results from section~5 shows that orbit--orbit operator and QED 
corrections may be of comparable size in light elements, though the 
orbit--orbit operators is of order ${\cal O}(\alpha^2)$ and QED 
corrections ${\cal O}(\alpha^3)$. It related with the fact that 
on the one hand a big number diagonal matrix elements of orbit--orbit 
operator are zeroes, 
%for Li--like sequence, 
on the other hand
off--diagonal matrix elements of orbit--orbit operator is of order
${\cal O}(\alpha^4)$. So, need to take into account bouth
orbit--orbit operator and QED corrections for studing the Li--like 
sequence and other light elements.

%From these studies, we conclude that the orbit-orbit operator is not an
%important effect on line-strengths computed in the length form at the 0.05\%
%level. In the case of energy levels, though the orbit-orbit operators is of
%order ${\cal O}(\alpha^2)$ and QED corrections ${\cal O}(\alpha^3)$, for
%different terms the corrections may be of comparable size in light elements.
%For the Li-like sequence, results are not improved by including one without
%the other.

%The results from section~5 shows that need to take into account 
%bouth orbit--orbit operatorand QED corrections for studing 
%the Li--like sequence.

\newpage

%"Orbit--orbit s\c aveikos operatoriaus bendrosios formos
%kai kurie nauji supaprastinimai."
"Supaprastintos orbit--orbit s\c aveikos operatoriaus formos taikymas 
lengviems atomams."

G. Gaigalas

Santrauka

Straipsnyje pateikta
bendra orbit--orbit s\c {a}veikos operatoriaus tenzorin\. {e}
forma. Ji susideda i\v s keturi\c u skirting\c u nari\c u $H_{oo1}^{(kk0,000)}$, 
$H_{oo2}^{(kk0,000)}$, $H_{oo3}^{(kk0,000)}$ ir $H_{oo4}^{(kk0,000)}$, turin\v ci\c u
t\c a pa\v ci\c a tenzorin\c e strukt\= ur\c a, ta\v ciau skirtingas radialiasias
dalis. Autorius surado tokius atvejus, kai darbe nagrin\. ejamas 
orbit--orbit s\c aveikos operatorius i\v ssirei\v skia per vien\c a nar\c i
$H_{oo3}^{(kk0,000)}$ t.y.
\c igyja \v zymiai paprastesn\c e i\v srai\v sk\c a nei buvo \v zinoma iki \v siol.
Straipsnyje pasi\= ulytas b\= udas, kaip \v sias naujas 
i\v srai\v skas b\= ut\c u galima naudoti bendrai, t.y. nepriklausomai nuo to tarp koki\c u
konfig\= uracij\c u ie\v skomi \v sio operatoriaus matriciniai elementai.
Tai leid\v zia: i) tiek diagonalius
tiek nediagonalius (konfig\= {u}racij\c {u} at\v {z}vilgiu) matricinius elementus
nagrin\. eti vieningai,
ii) efektyviau algoritmizuoti orbit--orbit s\c aveikos operatoriaus matricini\c u
element\c u skai\v ciavim\c a, iii) atlikti radiali\c uj\c u integral\c u tikslumo
\c ivertinim\c a.

Remiantis straipsnyje pasi\= ulyto metodo pagrindu atlikti
teoriniai Li izoelektronin\. {e}s sekos pagrindin\. {e}s b\= {u}senos ir
ionizacijos potencialo skai\v ciavimai. \v Siuo pasirinktu atveju, 
atsiranda nema\v zas skai\v cius toki\c u matricini\c u element\c u, kuriuos
nagrin\. ejant naudojamos supaprastintos orbit--orbit operatoriaus i\v srai\v skos.
Darbe gauti teoriniai rezultatai sutapo su teoriniais kit\c u autori\c u
rezultatais. Tai \c itikinamai parodo, jog darbe gautos supaprastintos
i\v srai\v skos bei pasi\= ulyta metodika (kaip jas efektyviai i\v snaudoti) yra
teisinga.

Nagrin\. ejant sud\. etingesnius atomus bei jonus, orbit--orbit operatorius ne visada
turi supaprastint\c a pavidal\c a. Tuo atveju \v s\c i operatori\c u tikslinga nagrin\. eti
kompleksi\v skai t.y. kur \c imanoma naudotis supaprastintoms, kur ne bendrosiomis
i\v srai\v skomis. \v Si problema taip pat straipsnyje i\v sspr\c esta.

\end{document}